\documentclass[fleqn,10pt]{wlscirep}

\usepackage{graphicx}
\usepackage{dcolumn}
\usepackage{bm}
\usepackage{color}
\usepackage{cancel}
\usepackage[normalem]{ ulem }
\usepackage{soul}
\usepackage{subfigure}

\graphicspath{{images/}{plots/}}

\newcommand{\avg}[1]{\ensuremath{\left< #1 \right>}}

\newcommand{\abs}[1]{\ensuremath{\left\vert#1\right\vert}}

\DeclareMathOperator{\Var}{Var}

\title{When open mindedness hinders consensus}
\author[1,*]{Hendrik Schawe}
\author[1,+]{Laura Hern\'{a}ndez}

\affil[1]{Laboratoire de Physique Th\'{e}orique et Mod\'{e}lisation, UMR-8089 CNRS, CY Cergy Paris Universit\'{e}, France}

\affil[*]{hendrik.schawe@u-cergy.fr}
\affil[+]{laura.hernandez@u-cergy.fr}

\begin{abstract}
    We perform a detailed study of the Hegselmann-Krause bounded confidence opinion dynamics model with
    heterogeneous confidence $\varepsilon_i$ drawn from uniform
    distributions in different intervals $[\varepsilon_l, \varepsilon_u]$.
    The phase diagram reveals a highly complex and nonmonotonous behaviour,
    with a re-entrant consensus phase in the region where fragmentation
    into multiple distinct opinions is expected for the homogeneous case.
    A careful exploration of the phase diagram, along with an extensive
    finite-size analysis, allows us to identify the mechanism leading to this
    counter-intuitive behaviour.
    This systematic study over system sizes which go well beyond those of
    previous works, is enabled by an efficient algorithm presented in this
    article.
\end{abstract}

\begin{document}
    % directly from the example
    \flushbottom
    \maketitle
    \thispagestyle{empty}

    \section{Introduction}
        Opinion formation and its dissemination within a society, a recurrent
        subject of study in social sciences, is often addressed through
        statistical analyses of data collected by the means of surveys or polls.
        By following the evolution of the statistical outcomes over time, it is
        possible to obtain some information about the dynamical processes underlying
        these phenomena. This broadly applied approach implicitly assumes a full
        mixing of the population, neglecting the role of the structure of
        interactions among social actors. However, the key role of such
        interactions had already been acknowledged long ago by E.~Durkheim~\cite{Durkheim},
        who coined the notion of \emph{social facts}. A social fact is a
        property characterizing the whole society instead of the individuals,
        which emerges as an outcome of the dynamics governed by the
        interactions among them. Very early works have collected data about
        these interactions in very small societies, using graph theory to
        represent them. J.~Scott~\cite{scott} gives a nice historical overview of
        network development in social sciences.
        Nowadays, the development of mathematical models along with the
        continuous growth of data collection on human activities proliferated
        in particular due to  the rise of online social networks, allow us to investigate different
        dynamical processes of opinion dynamics.

        The rationale behind  most opinion dynamics models is grounded in
        \emph{Social Influence Theory}~\cite{Kelman, Deutsch_Gerard} which
        assumes that as agents interact they may influence each other, making
        their opinions more alike. Accordingly, the dynamical rules that govern
        the evolution of the opinion of each agent, are often based on
        functions which aggregate the opinions of a given set of actors.
        Recently a complete mathematical classification of the possible
        outcomes of the dynamics based on aggregation functions in the case
        of discrete opinion variables has been obtained~\cite{Agnieszka,Poindron}.

        The mathematical description of the structure of social interactions
        allows to identify the neighbourhood of a given agent (the set of other
        agents directly connected to it) and it is now clear that this
        structure is relevant. Different neighbourhood choices along with the
        type of influence they have on each social actor have been considered,
        for example the adoption of the opinion of a randomly chosen neighbour \cite{voter},
        the adoption of the neighbours' majoritarian opinion \cite{majority}, or
        of agents whose opinion on other topics are already
        near the one of the target agent~\cite{axelrod}.

        Among all the studied models~\cite{Castellano2009Statistical,Sirbu2017Opinion},
        those representing the agents' opinion by a continuous variable are
        well suited to describe situations where the opinion on a particular
        problem is gradually built through the exchanges among the social
        actors. In particular \emph{bounded confidence} models consider that
        each agent will only interact with those agents whose opinions are
        already close to theirs and will not interact with the others.
        Noteworthy, as agents' opinions evolve with time, the network of social
        ties is not fixed.

        The best studied bounded confidence models are the \emph{Deffuant-Weisbuch} (DW)
        model \cite{deffuant2000mixing} and the \emph{Hegselmann-Krause} (HK)
        model \cite{hegselmann2002opinion}. In both models the opinion of
        agent $i$ is coded in a continuous variable and each agent may be
        influenced only by others whose opinion differs from his at most by a quantity called \emph{confidence}
        $\varepsilon_i$. Unlike the DW model, which considers pairwise
        interactions, in the HK model agents are synchronously influenced by
        all others within their confidence range.

        The most studied variant of the latter is the case of homogeneous confidence
        $\varepsilon_i = \varepsilon$, where large populations will always
        converge towards a uniform opinion, if the confidence is above a
        threshold $\varepsilon_c \approx 0.2$ \cite{hegselmann2002opinion,hegselmann2005opinion_eco}.

        However, a society is not a homogeneous collection of individuals. Some of them
        are \emph{open minded} and this property may be modelled by relatively
        large values of their confidence $\varepsilon_i$, accordingly the
        \emph{closed minded} ones will have low values of their confidence $\varepsilon_i$.
        Therefore heterogeneity of the confidence of single agents
        is an obvious and well motivated addition to the model.
        Until now heterogeneity in the HK model was mainly studied
        by introducing multiple subpopulations, each of them characterized by a specific confidence value, and  formed by a homogeneous set of agents.~\cite{Lorenz2010Heterogeneous,Kou2012multi,Pineda2015mass,Han2019Opinion}.
        Some other works study systems with random confidence, with
        $\varepsilon_i$ drawn from some distribution \cite{lorenz2003opinion,Liang2013Opinion,Shang2014agent},
        or from different distributions for multiple subpopulations \cite{Guiyuan2015Opinion}.
        Those studies are usually performed by the means of multi-agent
        simulations on small statistical samplings (50-100 samples),
        where each realization represents also a very small system (typically a few hundred agents).
        An alternative method, the multiple chain Markov model\cite{Lorenz2010Heterogeneous},
        has been proposed in order to obtain the properties of the infinite
        system.
        However, there, opinions are discretised which is known to
        to cause deviations from the HK model \cite{Fortunato2004Krause}
        and the probabilities of each opinion are derived from an initial set of agents.
        % However the probabilities of each opinion are still derived from
        % an initial small set of agents.
        Nevertheless, these works seemed to
        indicate that heterogeneity indeed leads to some surprising nonmonotonous
        effects in the dynamical outcomes. In particular, the existence of
        consensus in regions where the confidence of the agents is
        below the critical threshold of the homogeneous case have been
        reported~\cite{Pineda2015mass}.

        In this article, we present a systematic study of the phase diagram of
        the HK heterogeneous model in the parameter space given by the possible
        lower and upper bounds of the confidence values of the agents,
        $(\varepsilon_l, \varepsilon_u)$. We carefully sample the parameter
        space for large systems and we are able to obtain very good statistics
        for the studied quantities. Furthermore, we study the finite-size
        effects on the dynamics of the model, which reveals complex details of
        the consensus landscape that only become apparent
        at system sizes that are larger than those studied before. A careful
        study of the dynamical trajectories in the opinion space allows us to
        explain the re-entrant phase observed in the phase diagram.
        Finally, we introduce an algorithm that allows us to study system sizes
        and sampling statistics that go well beyond the present state of the art.

    \section{Models and Methods}
        We study the Hegselmann-Krause model (HK), which describes a compromise dynamic
        under bounded confidence. Each of $n$ \emph{agents} $i$ is endowed of a continuous variable   $x_i(t)$ representing \emph{opinion} and a
        fixed \emph{confidence} $\varepsilon_i$, modeling the heterogeneous idiosyncrasies of the agents. The \emph{neighbours} of agent $i$ are
        all agents $j$ with opinions inside the interval $[x_i - \varepsilon_i, x_i + \varepsilon_i]$, i.e.
        \begin{align}
            \label{eq:neighbours}
            I(i, \vec{x}) = \left\{ 1 \le j \le n | \abs{x_i - x_j} \le \varepsilon_i \right\}.
        \end{align}
        Note that every agent is a neighbour of itself.
        In each time step an agent $i$ talks to all its neighbours $j$ and adopts
        the average opinion of all neighbours, i.e.
        \begin{align}
            \label{eq:update}
            x_i(t+1) = \abs{I(i, \vec{x}(t))}^{-1} \sum_{j\in I(i, \vec{x}(t))} x_j(t).
        \end{align}

        This update is performed synchronously for all agents, although
         a sequential random update is possible. The latter typically leads  to longer
        convergence times, as single agents are left behind and may, for small
        values of $\varepsilon$, persist as isolated clusters in the final
        state.
        However, besides these effects, the observations which we describe
        in the following sections are qualitatively robust against the update schedule.

        The dynamic leads to a stable configuration, where the agents
        are converged to one or several opinions. The groups of agents ending in
        the same opinion are usually called \emph{clusters}. The situation in which
        only one giant cluster exists is called \emph{consensus}, two clusters are called
        \emph{polarization} and more are called \emph{fragmentation}. For the
        homogeneous case it is well known that above a critical $\varepsilon_c \approx 0.2$ \cite{Castellano2009Statistical}
        large systems will always converge to consensus and never reach consensus
        below this threshold.

        % \todo{extension with inhomogeneous confidence}
        While the original HK model uses homogeneous confidence $\varepsilon_i = \varepsilon$,
        the only modification we apply to this model, which will lead to
        surprisingly complex behaviour as we will see later, is to draw
        heterogeneous $\varepsilon_i$ from an i.i.d.\ uniform distribution bounded
        by two parameters $\varepsilon_l, \varepsilon_u$ with
        $\varepsilon_l \le \varepsilon_u$. The choice of the
        parameter $\varepsilon_l \ge 0$ determines how closed minded the most
        closed minded agents are and $\varepsilon_u \le 1$ determines how open
        minded the most open minded agents are.

        We are mainly interested in the effects of heterogeneity on consensus
        for the whole $(\varepsilon_l, \varepsilon_u)$ space.
        To evaluate this, we look at the relative size of the largest cluster $S$
        averaged over all simulated realizations $\avg{S}$. A value near $1$
        indicates consensus, smaller values indicate polarization or fragmentation.

        % \todo{b-tree based updating scheme}
        One difficulty in simulating the HK dynamics is that the run time scales
        in the number of agents as $\mathcal{O}(n^2)$, such that simulation of
        large system sizes becomes infeasible quickly. There are attempts to
        invent algorithms which are faster, but they come with trade offs, such
        as not simulating the actual dynamic, but generating an approximation for
        the $n\to\infty$ case \cite{Lorenz2007continuous} or necessitating a
        discretization of the model \cite{Fortunato2004Krause}.

        We will here introduce an algorithm, which is much faster than the
        naive approach for typical
        realizations, while preserving the continuous character down to the
        precision of the data types used. This enables us to simulate larger
        sizes with far better statistics than other contemporary studies of
        the HK model. Its fundamental idea is that to update agent $i$ we do
        only have to look at the agents within its confidence interval
        $[x_i-\varepsilon_i, x_i+\varepsilon_i]$, which
        are typically far fewer than $n$ for small
        values of $\varepsilon_i$ -- but in the order of $\mathcal{O}(n)$
        in the worst case.

        To achieve this in an efficient way, we maintain a
        \emph{binary search tree} (BST) \cite[p.~286]{cormen2009introduction}
        of all opinions $x_j(t)$\footnote{Technically, we use a B-tree \cite[p.~484]{cormen2009introduction}, which is a
        self-balancing generalization of a BST and better suited
        to iterate over contiguous ranges of entries.}. We use
        this datastructure to efficiently find the agent with the smallest
        opinion in the range $[x_i-\varepsilon_i, x_i+\varepsilon_i]$ and
        traverse the tree in order until we find the first node outside of this
        range, to calculate $x_i(t+1)$.

        Since BSTs can only store unique elements, we have to handle the case of
        multiple agents having the same opinion within precision of the used
        datatype. Therefore, we save at each node of the tree, additionally to
        the opinion, a \emph{counter}, which keeps track of the
        number of agents which have this opinion and are therefore represented
        by this node. Correspondingly this counter
        has to be handled as a weight when calculating the average.

        When the opinion of an agent is updated, it also has to be updated in
        the tree by removing its former opinion (or decreasing its counter by one)
        and inserting its current opinion (or increasing its counter by one).
        Both operations can be performed in time $\mathcal{O}(\log(n))$.
        This algorithm therefore has a typical time
        complexity for one step between $\Omega(n\log(n))$ in the best case and
        $\mathcal{O}(n^2)$ in the worst case.

        Note that the performance of this algorithm benefits from two
        independent effects.
        First, for agents with small confidence $\varepsilon_i$ we profit from
        the reduced number of opinions we need to look at to determine the
        average opinion of its neighbours.
        Second, the HK model tends to form clusters (within the chosen
        numerical precision) quickly, especially for large
        $\varepsilon_i$. Since clusters are represented
        in the BST as a single node with a high counter, the
        computational cost to calculate the average opinion is greatly reduced.

    \section{Results}
        We simulated the system for a broad range of system sizes $64 \le n \le 131072$
        and we carefully explored the parameter space of confidence intervals.
        For each run, we have drawn the confidence parameter of each
        agent $\varepsilon_i$ uniformly from an interval bounded by
        $(\varepsilon_l, \varepsilon_u) \in [0, 0.35] \times [0, 1]$.
        The simulations are run until the opinions, represented with 32 bit
        IEEE 754 float datatypes, converge.
        The convergence criterion we use here requires that the sum of the
        changes over all the agents is below a threshold,
        i.e.~$\sum_{i=1}^n \abs{x_i(t-1) - x_i(t)} < 10^{-4}$.

        Clusters are composed of agents holding the same opinion within a
        tolerance range of $10^{-4}$. We have checked that the results are
        robust when using different clustering criteria, e.g.~binning the
        opinion space.

\begin{figure}[htb]
            \centering

            % \includegraphics[scale=1]{map_16384}
            % \includegraphics[scale=1]{time_16384}

            % scirep wants single figure files
            \includegraphics[scale=1]{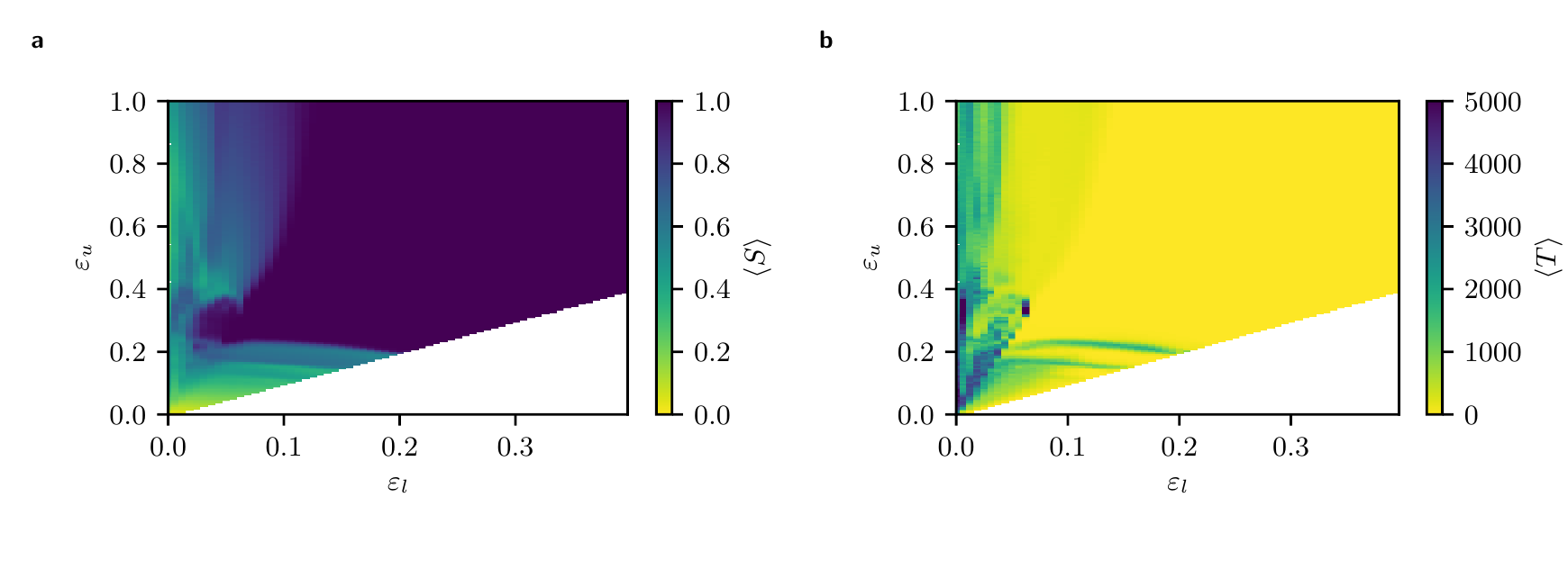}

            \caption{\label{fig:clustersize_maps}
                Left: Average relative size of the largest cluster $\avg{S}$.
                Right: Average convergence time $\avg{T}$.
                These data are collected for
                $8224$ pairs of $(\varepsilon_l, \varepsilon_u)$ for a system
                of $n=16384$ agents and averaged over $1000$ realizations.
                Note that three parameter pairs at $\varepsilon_l = 0 $ did not
                converge in reasonable computing time for all realizations.
                To avoid selection bias, they are therefore omitted and marked
                in white.
            }
        \end{figure}

        Figure~\ref{fig:clustersize_maps} shows a heat map of the average
        size of the largest cluster $\avg{S}$ for each of the $8224$ points
        that we have simulated in the  parameter space. The results shown by
        the heat map at each point $(\varepsilon_l, \varepsilon_u)$
        correspond to the average over 1000 simulated samples, which differ in
        their initial conditions although all represent the same society, where
        the heterogeneous confidence parameters of the agents $\varepsilon_i$
        have been uniformly drawn from the same interval.
        Note that the aspect ratio is not unity.

        The region not shown here ($[0.35, 1] \times [0, 1]$) corresponds to
        values of confidence intervals that always lead to consensus. This is
        expected, as both bounds are far above the critical value of the
        homogeneous case. The white triangle in the lower right corner
        consists of impossible intervals, where the lower bound would be higher
        than the upper one. The diagonal elements correspond to the homogeneous case,
        i.e.~$\varepsilon_l = \varepsilon_u$. One can see that on this diagonal,
        $\avg{S}$ changes from $1$ to approximately $1/2$ around
        $\varepsilon_l = \varepsilon_u = 0.2$, showing the transition from
        consensus to polarization that has been found in the homogeneous HK model \cite{Castellano2009Statistical}.

        The interesting results are situated on the left of the map, where a
        \emph{re-entrant consensus region}
        occurs around the point  $(0.05, 0.3)$. This reveals a nonmonotonous
        behaviour of the system as the fraction of confident agents increases
        in the population. Let us examine the map by considering a fixed value
        of $\varepsilon_l = 0.05$ while varying the upper value
        $\varepsilon_u$. We start at the homogeneous case $\varepsilon_u = 0.05$,
        where we find fragmentation. As $\varepsilon_u$ increases, a dark region is encountered
        showing strong consensus. This is expected as more agents with an increasing
        confidence (having a larger number of neighbours) enter the
        system, contributing to integrate agents with low confidence into the
        consensus group. However, as $\varepsilon_u$ increases further,
        consensus surprisingly disappears although the fraction of confident agents and the
        magnitude of their confidence is even larger.

        \begin{figure}[htb]
            \centering

            % \includegraphics[scale=1]{sweep}
            % \includegraphics[scale=1]{sweep_var}

            % scirep wants single figure files
            \includegraphics[scale=1]{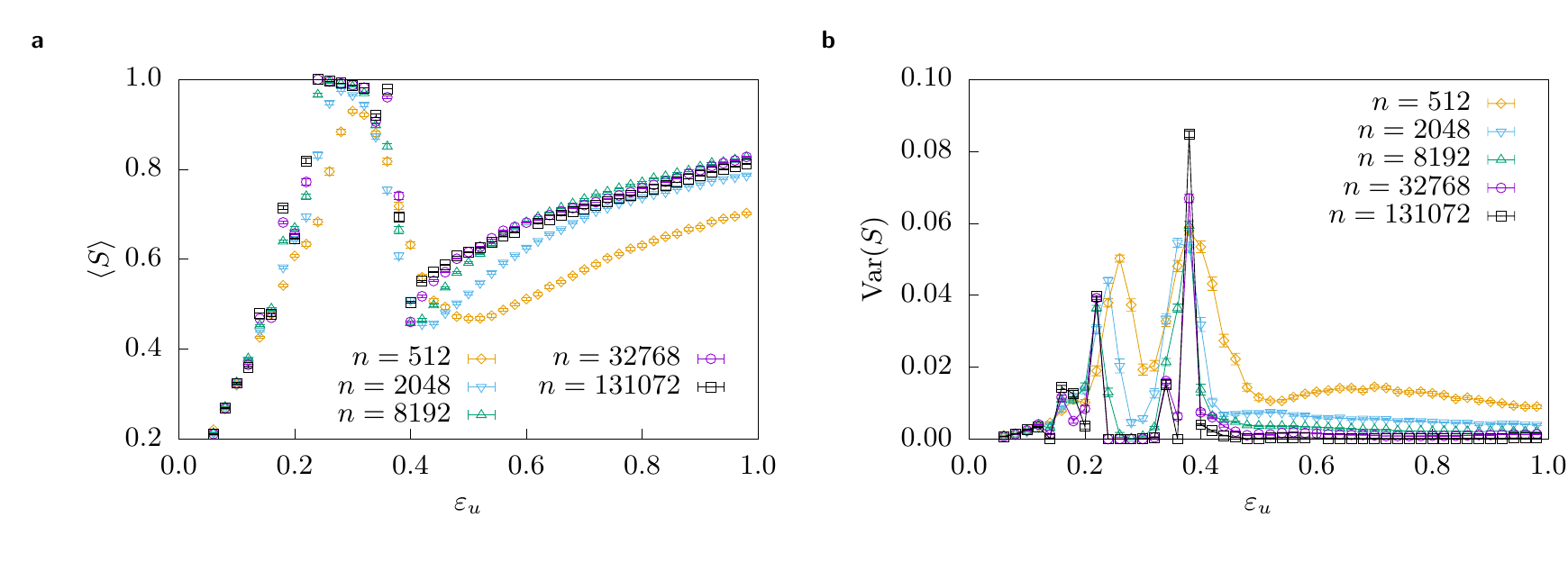}

            \caption{\label{fig:svseps:005}
                Left: Average size of the largest cluster $\avg{S}$.
                Fixed $\varepsilon_l = 0.05$ for varying values of
                $\varepsilon_u \in [\varepsilon_l, 1]$; this corresponds to a
                straight vertical line in Fig.~\ref{fig:clustersize_maps}
                through the re-entrant consensus phase. The consensus region is robust
                with increasing system sizes. Also note that the behaviour is
                highly complex as the behaviour with increasing system size is
                not always monotonous, e.g.~at $\varepsilon_u = 0.2$ or
                $\varepsilon_u = 0.42$.
                Right: Variance of the size of the largest cluster $\Var(S)$.
                The transition points into consensus and out of consensus
                can be located at the peaks of the variance. Using the
                finite-size behaviour, we estimate them for $\varepsilon_l = 0.05$ at
                $\varepsilon_u = 0.22(1)$ and $\varepsilon_u = 0.38(1)$.
                Lines are just guides to the eye.
            }
        \end{figure}

        The left panel of Fig.~\ref{fig:svseps:005} shows the evolution of the order parameter
        $\avg{S}$ with $\varepsilon_u$ for a vertical cut in the region of the
        re-entrant phase of the heatmap shown in Fig.~\ref{fig:clustersize_maps}
        and described in the previous paragraph in higher detail and for
        different system sizes. For a system that contains very
        closed minded agents, when the fraction of open minded
        agents increases, the size of the largest cluster also increases leading
        to consensus, however as the fraction of confident agents increases
        further, consensus is lost.

        The right of Fig.~\ref{fig:svseps:005} shows the fluctuation of the
        order parameter $\avg{S}$. There are two peaks visible, which indicate
        the location of the two transitions, in and out of the re-entrant phase.
        Interestingly, the finite-size behaviour signals a real double transition
        as the peaks increase and separate with increasing size, suggesting that
        this is not just a finite-size effect, but that it remains in the
        thermodynamic limit, i.e.~the $n \to \infty$ limit. Using these data,
        we estimate the re-entrant phase to span the
        interval $\varepsilon_u \in [0.22(1), 0.32(1)]$ for fixed $\varepsilon_l = 0.05$.

\begin{figure*}[htb]
            \centering

            \includegraphics[scale=1]{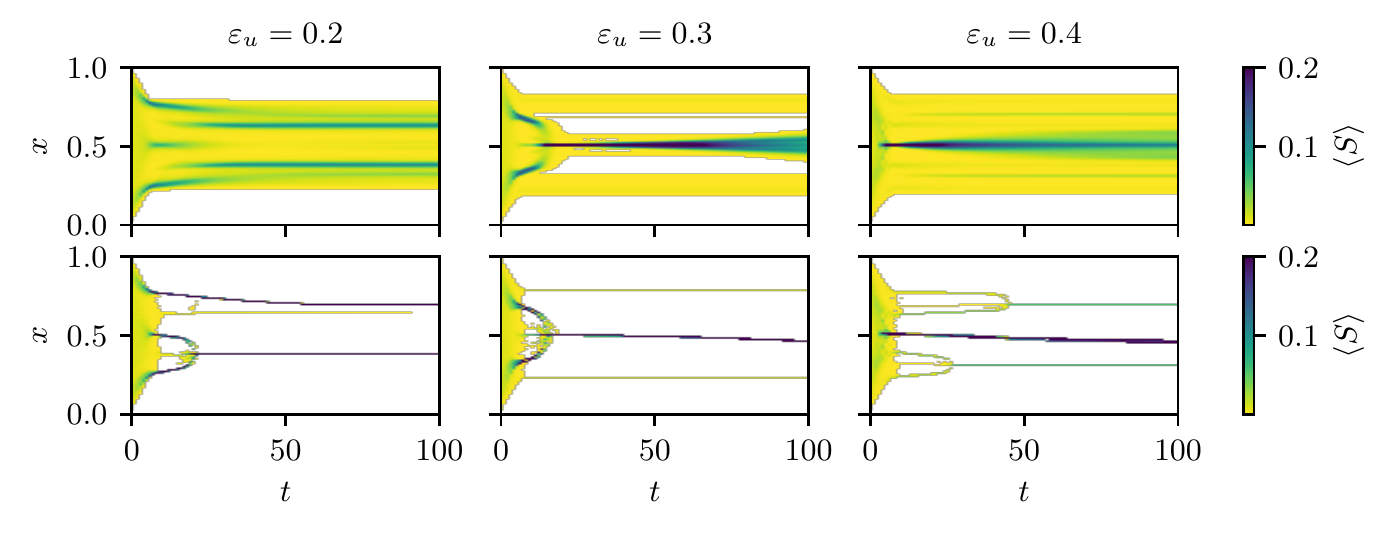}

            \caption{\label{fig:timedevelopment}
                We look for fixed $\varepsilon_l = 0.05$ and $n=16384$ at three values of
                $\varepsilon_u \in \{0.2, 0.3, 0.4\}$ (from left to right) corresponding to one
                value left of the peak, one at the peak
                and one right of the peak of Fig.~\ref{fig:svseps:005}.
                The images show which fraction of agents have a given opinion
                $x$ for each time step.
                % We prefer this visualization instead of the popular one,
                % where single trajectories are traced by lines. The latter
                % approach becomes impractical for large numbers of agents
                % and cannot express the size of clusters.
                Note that the colour scale is truncated
                at $0.3$ to generate images with a good contrast in the
                interesting region at small times, such that the darkest colour
                can also represent values larger than $0.3$.
                The time axis is also truncated to focus on the most
                interesting region of small times, i.e.~not the whole range
                until reaching a stationary state is visualized.
                Top: aggregated statistics over $10000$ realizations of the
                initial conditions.
                Bottom: single trajectories.
            }
        \end{figure*}

        In order to explain this paradoxical result, a careful examination of
        the dynamic behaviour of the agents' opinions is necessary. In
        Fig.~\ref{fig:timedevelopment} we show the evolution of opinions for
        systems corresponding to three different values of $\varepsilon_u$ in
        the re-entrant region. The top row corresponds to averages
        over $10000$ realizations of the initial conditions, while the bottom
        row shows, as examples, the evolution of opinions of a single realization
        each.

        The mechanism leading to the observed behaviour is rooted in the
        different characteristic times that open and closed minded agents need
        to join a majoritarian opinion strand.
        The latter take more time to reach the consensus opinion as they need
        to meet agents within their narrow confidence interval. This is at the
        origin of the bell shaped structures in the bottom left panel of Fig.~\ref{fig:timedevelopment}

        These structures enable closed minded agents from the whole region over
        which the bell spans, to join a strand and evolve.
        If the strand counts enough open minded agents, they can pull it  towards a consensus opinion, bringing with them the closed minded
        agents that have an opinion close to theirs. This is what happens in the
        bottom middle panel of Fig.~\ref{fig:timedevelopment} at $\varepsilon_u = 0.3$.
        Also the signature of this structure is clearly visible in the average
        over many realization in the top middle panel.
        The two smaller strands visible in this panel contain only 3\% of the
        agents. As it is very frequent that the central cluster converges to
        one of them, the averaged picture shows an apparent diffusion of the
        central opinion for long times. Moreover we expect the small clusters to
        vanish in the thermodynamic limit, since the peak of Fig.~\ref{fig:svseps:005}a
        shows an upward trend for increasing system sizes.

        Note that agent $i$ can in each time step move at most $\varepsilon_i$,
        such that closed minded agents need to see other agents for at least a
        few iterations to be able to change their opinion from one extreme to the
        consensus opinion at $0.5$.
        More open minded agents (like those that appear when increasing $\varepsilon_u$)
        will evolve very quickly to a central opinion, because
        they can interact with a large fraction of the other agents and therefore
        are able to jump directly into the centre. As a consequence closed
        minded agents are left behind and are not able to join the consensus
        opinion.
        This is the situation depicted in the right panels of Fig.~\ref{fig:timedevelopment}.

        The case of $\varepsilon_u = 0.4$, shown on the right of Fig.~\ref{fig:timedevelopment},
        seems to show that the central strand contains about 90\%, which is an
        apparent contradiction with the mean cluster size $\avg{S} \approx 0.5$
        shown in Fig.~\ref{fig:clustersize_maps}. The solution to this discrepancy
        is that the central strand located at an opinion $x \approx 0.44 $ splits
        in the stationary state to three very close clusters, therefore there are 5 clusters
        in total, which are composed of different groups of agents. One of them
        contains about 50\% of the agents, mainly having large confidence intervals,
        able to interact with agents of the small strands located at $x \approx 0.32$
        and $x \approx 0.68$. Another cluster contains about 30\% of the agents,
        which can interact only with the bottom strand at $x \approx 0.32$ (and
        all agents of the central strand), but not with the upper one at $x \approx 0.68$, such that
        it will converge to a slightly lower opinion. The last cluster contains
        about 10\% of the agents which can only interact with those in the
        central strand (their opinion being in the middle of the other two
        clusters forming the central violet strand). Therefore, the size of the
        largest cluster of this realization is $S \approx 0.5$, the
        value observed in the valley of Fig.~\ref{fig:svseps:005}.
        Note that this effect can easily be missed when using a more discrete
        cluster criterion, like binning with too few bins.

        Figure~\ref{fig:clustersize_maps}b confirms these results: average
        convergence time in the region of re-entrant consensus is much larger
        than on neighbouring regions, illustrating the extra time needed by closed
        minded agents to join the consensus strand \emph{pulled} by
        open minded agents.

        Interestingly, the bell shapes observed in the evolution of opinions,
        which help to integrate isolated agents into a single strand, are
        observed for or all the parameters shown here. However,
        when $\varepsilon_u$ is either too low or too high, these structures
        are mainly formed by agents with very low confidence, in the first case
        just because they are the majority of the population and in the second
        because agents with large confidence have already joined the main
        strand. For intermediate values of $\varepsilon_u$, these bell
        structures contain both open and closed minded
        agents, and the former may bring the latter into the final consensus strand.

        %\todo{Delete this part, if the manuscript becomes too long. Limit are 4500 words ~ 5 pages (excluding literature and abstract), also see \url{https://journals.aps.org/pre/edannounce/10.1103/PhysRevE.90.060001}}

\begin{figure}[htb]
            \centering
            \includegraphics[scale=1]{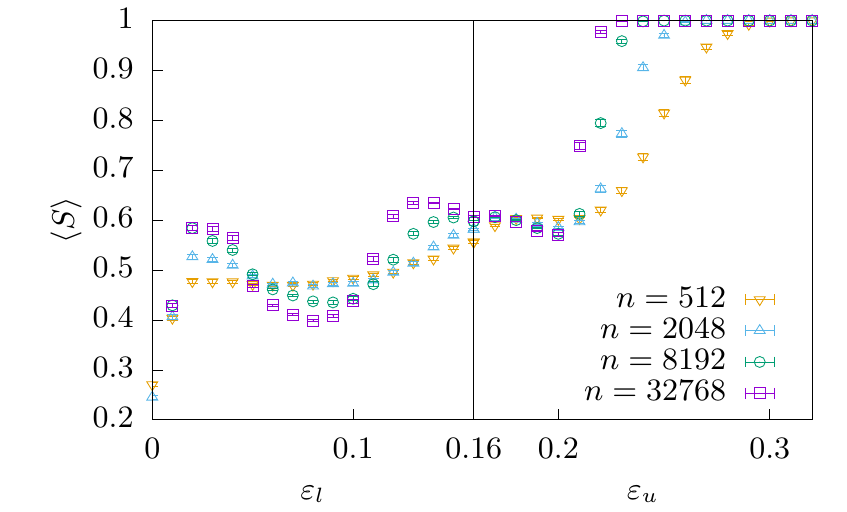}
            \caption{\label{fig:svseps:016}
                Average size of the largest cluster $\avg{S}$.
                The left half shows $\varepsilon_u = 0.16$ fixed and
                $\varepsilon_l$ varying, the right half shows
                $\varepsilon_l = 0.16$ fixed and $\varepsilon_u$ varying.
                The position marked by the vertical line is the homegeneous
                case $\varepsilon_l = \varepsilon_u = 0.16$. This corresponds
                to a straight line in Fig.~\ref{fig:clustersize_maps} reflected
                at the diagonal.
            }
        \end{figure}

        Another peculiarity visible in Fig.~\ref{fig:clustersize_maps} is
        the extremely complex behaviour of $\avg{S}$. For
        example at $\varepsilon_u \approx 0.16$ there is a local maximum. This
        is reminiscent of an effect called ``consensus strikes back'' by
        Ref.~\cite{lorenz2006consensus} for a related ``interacting Markov chain''
        (and not agent-based) HK model. There, a consensus phase was
        observed for the homogeneous case for $0.152 \le \varepsilon \le 0.174$.
        We show finite size data in Fig.~\ref{fig:svseps:016}, where one bound
        (either $\varepsilon_l$ or $\varepsilon_u$) is fixed to $0.16$ while
        the other bound is indicated in the horizontal axis, to test whether
        we can see a signature of this effect in the agent-based HK model.
        Indeed, we can identify an upward trend in $\avg{S}$, which becomes even
        stronger with added inhomogeneities at and slightly below $0.16$.

        Although we do not attempt to extrapolate this to larger sizes, due
        to the nonmonotonous behaviour observed in this region of the parameter
        space, our results indicate that there is indeed a trend of increasing
        cluster size located approximately in the region where the
        ``consensus strikes back'' effect was observed in the homogeneous case
        for an related model \cite{lorenz2006consensus}. However, also with
        heterogenity, which we showed to be beneficial for consensus, the
        agent-based HK model does not show consensus at the system sizes
        we observed.

    \section{Conclusions}
        We have performed a very detailed characterisation of the phase diagram
        of the heterogeneous Hegselmann-Krause model by means of an efficient
        algorithm that allows the simulation of large samples. In this way we
        could obtain very good statistics and investigate finite-size effects
        overcoming the size limitations of previous works. Our results reveal
        a nonmonotonous behaviour with a re-entrant consensus phase in the
        region of the parameter space where fragmentation is expected.

        Previously, the phase diagram of the HK model or closely related models
        with two different values of the bounded confidence coexisting in the population has been studied,
        also revealing a nonmonotonous behaviour depending on the amount of
        open and closed minded agents~\cite{Lorenz2010Heterogeneous,Pineda2015mass}. Here, by the means
        of the simulation of large systems, we characterise the rich behaviour
        of the fully heterogeneous HK model, and we identify the region
        $[\varepsilon_l,\varepsilon_u]$, where the consensus phase clearly
        enters in the region, in which one would not expect consensus based on
        the behaviour of the homegeneous case.
        In particular, we find that increasing the proportion of open minded
        agents may lead to a loss of consensus, provided closed minded agents
        are still in the system. We were able to explain this counter-intuitive
        observation with a careful study of the opinion evolution. Its
        origin is the slow movement of closed minded agents in the opinion space.
        When the system contains a large fraction of very open minded agents,
        who converge very quickly to a majoritarian opinion, the closed minded
        agents are left behind and full consensus is precluded. On the contrary
        it is easier to reach consensus when the closed minded agents coexist
        with a fraction of moderately open minded one, who converge to the
        majoritarian opinion much slower. This relative slow convergence allows
        for the interaction with the closed minded agents who are slowly
        dragged to the majoritarian opinion.

        We believe that the introduction of quenched disorder by increasing
        the complexity of the HK model offers a lot of potential for further
        studies. Variations of the model that introduce  a
        network of social ties, which limits the possible neighbourhoods of any
        agent, or the integration of cost that the agents must
        bear for changing their opinion, are work in progress.

    \section*{Acknowledgments}
       This work was supported by the OpLaDyn grant obtained in the 4th round
       of the TransAtaltic program Digging into Data Challenge (2016-147 ANR OPLADYN TAP-DD2016)
       and Labex MME-DII (Grant No. ANR reference 11-LABX-0023).

    \bibliography{lit}

\end{document}